\documentclass{aa} 
\usepackage{hyperref}   
\hypersetup{colorlinks=true,linkcolor=[rgb]{1.,0.2,0.2},citecolor=[rgb]{0.1,0.4,1.},filecolor=[rgb]{0.7,0.2,0.2},urlcolor=[rgb]{0.7,0.2,0.2}}
\usepackage{pdflscape}

\usepackage{soul}
\usepackage{float}
\usepackage{color}
\usepackage{times}
\usepackage{natbib}
\usepackage{amsmath,amssymb,amsfonts,txfonts}
\usepackage[colorinlistoftodos]{todonotes}
\usepackage{graphicx}
\usepackage{txfonts}
\definecolor{mycolor}{rgb}{0.58, 0.0, 0.83}
\definecolor{darkspringgreen}{rgb}{0.09, 0.45, 0.27}

\begin{document}

   \title{Low-redshift measurement of the sound horizon \\ through gravitational time-delays}

   \author{Nikki Arendse\inst{1}\thanks{email: nikki.arendse@nbi.ku.dk ; adriano.agnello@nbi.ku.dk ; 
             radek.wojtak@nbi.ku.dk}
          \and
          Adriano Agnello\inst{1}
          \and
          Rados{\l}aw ~J. ~Wojtak\inst{1}
          }

   \institute{
             \inst{1} DARK, Niels Bohr Institute, University of Copenhagen, Lyngbyvej 2, 2100 Copenhagen, Denmark\\
             }

 \date{Received 28 May 2019 / Accepted 4 October 2019}


  \abstract
   {The matter sound horizon can be infered from the cosmic microwave background within the Standard Model. Independent direct measurements of the sound horizon are then a probe of possible deviations from the Standard Model. }
   {We aim at measuring the sound horizon $r_s$ from low-redshift indicators, which are completely independent of CMB inference. }
   {We used the measured product $H(z)r_s$ from baryon acoustic oscillations (BAO) together with supernovae~\textsc{I}a to constrain $H(z)/H_{0}$ and time-delay lenses analysed by the H0LiCOW collaboration to anchor cosmological distances ($\propto H_{0}^{-1}$). {Additionally, we investigated the influence of adding a sample of quasars with higher redshift with standardisable UV-Xray luminosity distances.} {We adopted polynomial expansions in $H(z)$ or in comoving distances} so that our inference was completely independent of any cosmological model on which the expansion history might be based. Our measurements are independent of Cepheids and systematics from peculiar motions {to within percent-level  accuracy.}}
   {The inferred sound horizon $r_s$ varies between $(133 \pm 8)$~Mpc  and $(138 \pm 5)$~Mpc across different models. The discrepancy with CMB measurements is robust against model choice. Statistical uncertainties are comparable to systematics.}
   {The combination of time-delay lenses, supernovae, and BAO yields a distance ladder that is independent of cosmology (and of Cepheid calibration) and a measurement of $r_s   $ that is independent of the CMB. These cosmographic measurements are then a competitive test of the Standard Model, regardless of the hypotheses on which the cosmology is based. }


   \maketitle

\section{Introduction}

The sound horizon is a fundamental scale that is set by the physics of the early Universe and is imprinted on the clustering of dark and luminous matter of the Universe. The most precise measurements of the sound horizon are obtained from observations of the acoustic peaks in the power spectrum of the cosmic microwave background (CMB) radiation, {although the inference partially depends on the underlying cosmological model}. In particular, the recent \textit{Planck} satellite mission yielded a sound horizon scale (at the end of the baryonic drag epoch) of $r_{\rm s}=147.09\pm0.26$~Mpc. This was based on the spatially flat six-parameter $\Lambda$CDM model, which provides a satisfactory fit to all measured properties of the CMB \citep{Planck2018}, and on the Standard Model of particle physics.

The sound horizon remains fixed in the comoving coordinates since the last scattering epoch and its signature can be observed at low redshifts as an enhanced clustering of galaxies. This feature is referred to as baryon acoustic oscillations (BAO). 
When we assume that the sound horizon is calibrated by the CMB, BAO observations can be used to measure distances and the Hubble parameter at the corresponding redshifts. The resulting BAO constraints can then be extrapolated to $z=0,$ for instance, using type~Ia supernovae (SNe), in order to determine the present-day expansion rate $H_0.$ However, this \textit{\textup{inverse distance ladder}} procedure depends on the choice of cosmological model and on the strong assumption that the current standard cosmological model provides an accurate and sufficient description of the Universe at the lowest and highest redshifts. The robustness of the standard cosmological model has recently been questioned on the grounds of a strong and unexplained discrepancy between the local $H_0$ measured from SNe with distances calibrated by Cepheids and its CMB-based counterpart \citep[currently a $4.4\sigma$ difference;][]{Riess2019}. The inverse distance ladder calibrated on the CMB should therefore be taken with caution. Recently, \citet{mac19} performed an inverse-distance-ladder measurement of $H_0$ adopting the baseline $r_s$ from \textit{Planck}, and therefore their inferred $H_0$ agrees with CMB predictions, as expected. 

Observations of BAO alone only constrain a combination of the sound horizon and a distance or the expansion rate at the corresponding redshift, that is,\footnote{Distances are defined more precisely below.} $r_{\rm s}/D(z)$ and $r_{\rm s}H(z).$ Using SNe, we can propagate BAO observables to redshift $z=0$ and obtain constraints on $r_{\rm s}H_{0}$ that are fully independent of the CMB \citep{Hui2017,Sha2018}. The extrapolation to low redshifts can be performed using various cosmographic techniques, so that the final measurement is essentially independent of cosmological model. Furthermore, combining BAO constraints with a low-redshift absolute calibration of distances or the expansion history, we can break the intrinsic degeneracy of the BAO between $r_{\rm s}$ and $H(z)$ and thus can determine the sound horizon scale. The resulting measurement is based solely on low-redshift observations, and it is therefore an alternative based on the local Universe to the sound horizon inferred from the CMB.

Several different calibrations of distances or the expansion history have been used to obtain independent low-redshift measurements of the sound horizon. The main results include the calibration of $H(z)$ estimated from cosmic chronometers \citep{Hea2014,Ver2017}, the local measurement of $H_0$ from SNe with distances calibrated with Cepheids \citep{Ber2016}, angular diameter distances to lens galaxies \citep{Jee2016, Woj2019}, and adopting the Hubble constant from time-delay measurements \citep{Ayl2019}, although the last measurement is based on cosmology-dependent modelling \citep{Bir2019}. Currently, the sound horizon is most precisely constrained by a combination of BAO measurements from the \textit{Baryon Oscillations Spectroscopic Survey} \citep[BOSS;][]{Alam2017}, with a calibration from the \textit{Supernovae and $H_{0}$ for the Equation of State of dark energy} project \cite[SH0ES; ][]{Riess2019}. A significantly higher local value of the Hubble constant than its CMB-inferred counterpart implies a substantially smaller sound horizon scale than its analogue inferred from the CMB under the assumption of the standard $\Lambda$CDM model \citep{Ayl2019}. The discrepancy in $H_0$ and $r_s$ may indicate a generic problem of distance scale at lowest and highest redshifts within the flat $\Lambda$CDM cosmological model \citep{Ber2016}. 

Here, we present a self-consistent inference of $H_0$ and $r_{\rm s}$ from BAO, SNe~\textsc{I}a, and time-delay likelihoods released by the H0LiCOW collaboration \citep{Suyu2010, Suyu2014, Wong2017, Suyu2017, Bir2019}. We examine flat-$\Lambda$CDM models as a benchmark and different classes of cosmology-free models. Our approach allows us to determine the local sound horizon scale in a model-independent manner. A similar method was employed by \cite{Taub2019}, who used SNe to extrapolate constraints from time-delays to redshift $z=0,$ and thus to obtain a direct measurement of the Hubble constant that depends rather weakly on the adopted cosmology.

This paper is organised as follows. The datasets, models, and inference are outlined in Section 2. Results are given in Section 3, and their implications are discussed in Section 4.

Throughout this work, comoving distances, luminosity distances, and angular diameter distances are denoted by $D_{M},$ $D_{L},$ and $D_{A}$, respectively. We also adopt the distance duality relations $D_{M}(z_{1}<z_{2})=D_{L}(z_{1}<z_{2})/(1+z_{2}),$ $D_{A}(z_{1}<z_{2})=D_{M}(z_{1}<z_{2})/(1+z_{2}),$
which should hold in all generality and whose validity with current datasets has been tested \citep{Woj2019}.

\vspace{-0.1cm}

\section{Datasets, models, and inference}
We used a combination of different low-redshift probes to set different distance measurements and different models for the expansion history. All models inferred the following set of parameters: $H_{0},$ $r_s,$ $M_1$ (normalisation of the SN distance moduli), and coefficients parametrising the expansion history or distance as a function of redshift. Curvature $\Omega_k$ is left as a free parameter in some models. The sample of high-redshift quasars introduces two additional free parameters: the normalisation $M_2$ and the intrinsic scatter $\sigma_{int}$ of the quasar distance moduli.

\subsection{Models}
The first model, for homogeneity with previous literature, adopted a polynomial expansion of $H(z)$ in $z$: 
\begin{equation}
H(z) = H_0 \; (1 + \mathcal{B}_1 z + \mathcal{B}_2 z^2 + \mathcal{B}_3 z^3 ) + \mathcal{O}(z^4)
\label{Hmodel1}
,\end{equation}
where the coefficients are related to the standard kinematical parameters, that is, the deceleration $q_{0}$, jerk $j_{0}$ , and snap $s_{0}$, in the following way \citep{Xu2011,Weinberg2008,Visser2004}:
\begin{align*}
&\mathcal{B}_1 = 1 + q_0 \\ 
&\mathcal{B}_2 = \frac{1}{2} (j_0 - q_0^2) \\
&\mathcal{B}_3 = \frac{1}{6} (3q_0^3 + 3q_0^2 - j_0 (3 + 4q_0) - s_0)\ . 
\end{align*}
Model distances were computed through direct integration of $1/H(z).$ This is preferred over a corresponding expansion in distances \citep[as chosen e.g. by][]{mac19} in order to ensure sub-percent accuracy in the model distances (Arendse et al., in prep.).

In our second chosen model family, $H(z)$ was expanded as a polynomial in $x=\log(1+z):$

\begin{equation}
H(x) = H_0 \; (1 + \mathcal{C}_1 x + \mathcal{C}_2 x^2 + \mathcal{C}_3 x^3) + \mathcal{O}(x^4).
\end{equation}
Plugging the Taylor expansion of $z = 10^x - 1$ into equation \ref{Hmodel1} and grouping the new terms by order, we find the following mapping between coefficients $C_{i}$ and the kinematical parameters:

\begin{align*}
&\mathcal{C}_1 = \ln(10) \; (1 + q_0) \\ 
&\mathcal{C}_2 = \frac{\ln^2(10)}{2} \; (- q_0^2 + q_0 + j_0 + 1) \\
&\mathcal{C}_3 = \frac{\ln^3(10)}{6} (3q_0^3 + q_0 (1 - 4 j_0) - s_0 + 1)\ .
\end{align*}
Here, distances were also computed through direct numerical integration of $1/H(z)$.

In our third model choice, comoving distances were computed through expansion in $y=z/(1+z),$ and $H(z)$ was obtained through a general relation \citep[][]{Li2019},
\begin{equation}
H(z, \Omega_k) = \frac{c}{\partial D_M(z)/\partial z} \; \sqrt{1 + \frac{H_0^2 \Omega_k}{c^2} D_M(z)^2}\ .
\end{equation}
When a polynomial expansion \begin{equation}
D_{M}(y)=\frac{\mathrm{c}}{H_{0}}\left(\ y+\mathcal{D}_{2}y^{2}/2+\mathcal{O}(y^{3})\ \right)
\end{equation}
is adopted, then the second-order coefficient $\mathcal{D}_2$ is related to the deceleration parameter $q_0$ through
\begin{equation}
q_{0}=1-\mathcal{D}_{2}\ .
\end{equation}

Adopting multiple families of parametrisations, for $H(z)$ and/or for model distances allowed us to quantify the systematics due to different ways of extrapolating the given distance measurements down to $z=0.$ This is equivalent to another common choice of adopting different cosmologies to extend the CDM model, but with the important difference that our chosen parametrisations are completely agnostic about what the underlying cosmological model should be.

Lastly, for the sake of comparison with widely adopted models, we also adopted a $\Lambda$CDM model class, with a uniform prior $\Omega_{k}=[-1.0,1.0]$ on curvature, and with the constraint that $\Omega_{\Lambda}+\Omega_{m}+\Omega_{k}=1.$ A discrepancy in flat-$\Lambda$CDM ($\Omega_{k}=0$) between CMB measurements and our low-redshift measurements would then indicate that more general model families are required, that is, possible departures from concordance cosmology, or that the Standard Model needs to be extended. 

\subsection{Datasets}

Our measurement relies on the complementarity of different cosmological probes. BAO observations constrain $r_{s}H(z)$ at several different redshifts and independently of the CMB. Standard candles play the role of the inverse distance ladder, by means of which the BAO constraints can be extrapolated to redshift $z=0.$ Finally, gravitational lensing time-delays place constraints on $H_{0},$ thus breaking the degeneracy between $H_{0}$ and $r_{s}$ in the inverse distance ladder of BAO and standardisable candles.

In our study, we used pre-reconstruction (independent of cosmological model) consensus measurements of the BAO from the Baryon Oscillations Spectroscopic Survey \citep{Alam2017}. For the relative luminosity distances, we employed binned distance moduli of SNe Ia from the Pantheon sample \citep{Scol2018}. We excluded possible changes due to the choice of SN sample by re-running our inference on distance moduli from JLA \citep{Bet2014}, and with the current quality of data, there is no appreciable change in the results. Finally, we used constraints on time-delays of four strongly lensed quasars observed by the H0LiCOW collaboration \citep[see][and references therein]{Suyu2017,Bir2019}.
Results from a fifth lens have recently been communicated by H0LiCOW \citep{Rusu2019}. We currently use only results that have been reviewed, validated, and released.

 \noindent As an option that provides more precise distance indicators at high redshifts, we used distance moduli estimated from a relation between UV and X-ray luminosity quasars, which was proved to be an alternative standard candle at high redshift \citep{Ris2018}. \citet{Ris2018} reported that quasar distances at high redshift show a deviation from $\Lambda$CDM; however, the lack of any corroborative pieces of evidence does not allow us to conclude if this deviation is a genuine cosmological {\color{white}anomaly}

\begin{figure}[H]
        \centering
        \includegraphics[width=\linewidth]{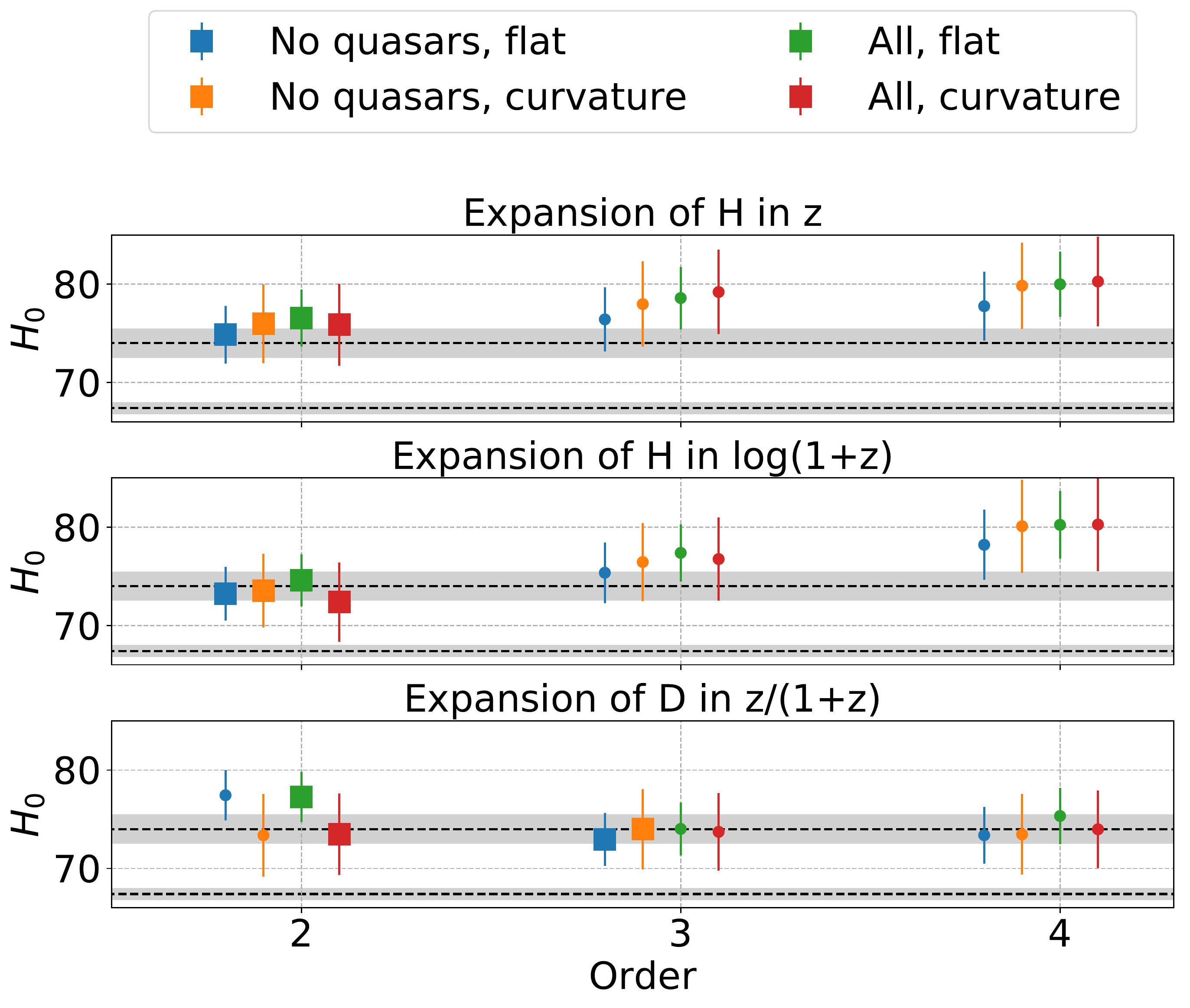}
        \caption{Inferred Hubble constant $H_0$ (in km/s/Mpc) vs. the chosen model family and expansion truncation. The fiducial values from each expansion model (displayed as squares) are chosen by considering the change in BIC score and in $\ln\mathcal{L}_{m.a.p.}$ vs. the change in degrees of freedom. The upper dashed line corresponds to the local measurement value of $H_{0} = 74.0$ km~s$^{-1}$~Mpc$^{-1}$ with a Cepheid calibration, and the lower dashed line corresponds to the Planck value of $H_{0} = 67.4$ km~s$^{-1}$~Mpc$^{-1}$. The shaded grey regions show the error bars.}
        \label{fig:trends}
\end{figure}

 \noindent anomaly or an unaccounted-for systematic effect. For this reason, we dismissed the quasar data at redshifts $z>1.8$, which is the highest redshift of lensed quasars in our sample.

\subsection{Inference}

The best-fit parameters and credibility ranges of the different expansion models were obtained by sampling the posterior using affine-invariant Monte Carlo Markov chains \citep{gw2010}, and in particular with the python module \texttt{emcee} \citep{fm13}. For the BAO and SN data set, the uncertainties are given by a covariance matrix \textbf{C}. The likelihood is obtained by
\begin{eqnarray}
\nonumber \mathcal{L}\ &=&\ p(\mathrm{data|model})\ \propto e^{- \chi^2 / 2}\ ,\\
\chi^2 &=& \textbf{r}^{\dagger} \textbf{C}^{-1} \textbf{r}
\end{eqnarray}
where \textbf{r} corresponds to the difference between the value predicted by the expansion and the observed data. 

The high-redshift quasar sample contains significant intrinsic scatter, $\sigma_{\textrm{int}},$  which has to be modelled as an additional free parameter.  The total uncertainty on each quasar data point is the sum of $\sigma_{i},$ the uncertainty of that data point, and $\sigma_{\textrm{int}}.$ This leads to the following formula of the likelihood:
\begin{equation}
\mathcal{L}_{\textrm{quasars}} = \sum_{i=1}^N \frac {\mathrm{e}^{ - r_i^2 / 2 (\sigma_i^2 + \sigma_{\textrm{int}}^2)} }{\sqrt{(\sigma_i^2 + \sigma_{\textrm{int}}^2) 2\pi }} \; 
.\end{equation}

\begin{figure}[b]
        \centering
        \includegraphics[width=\linewidth]{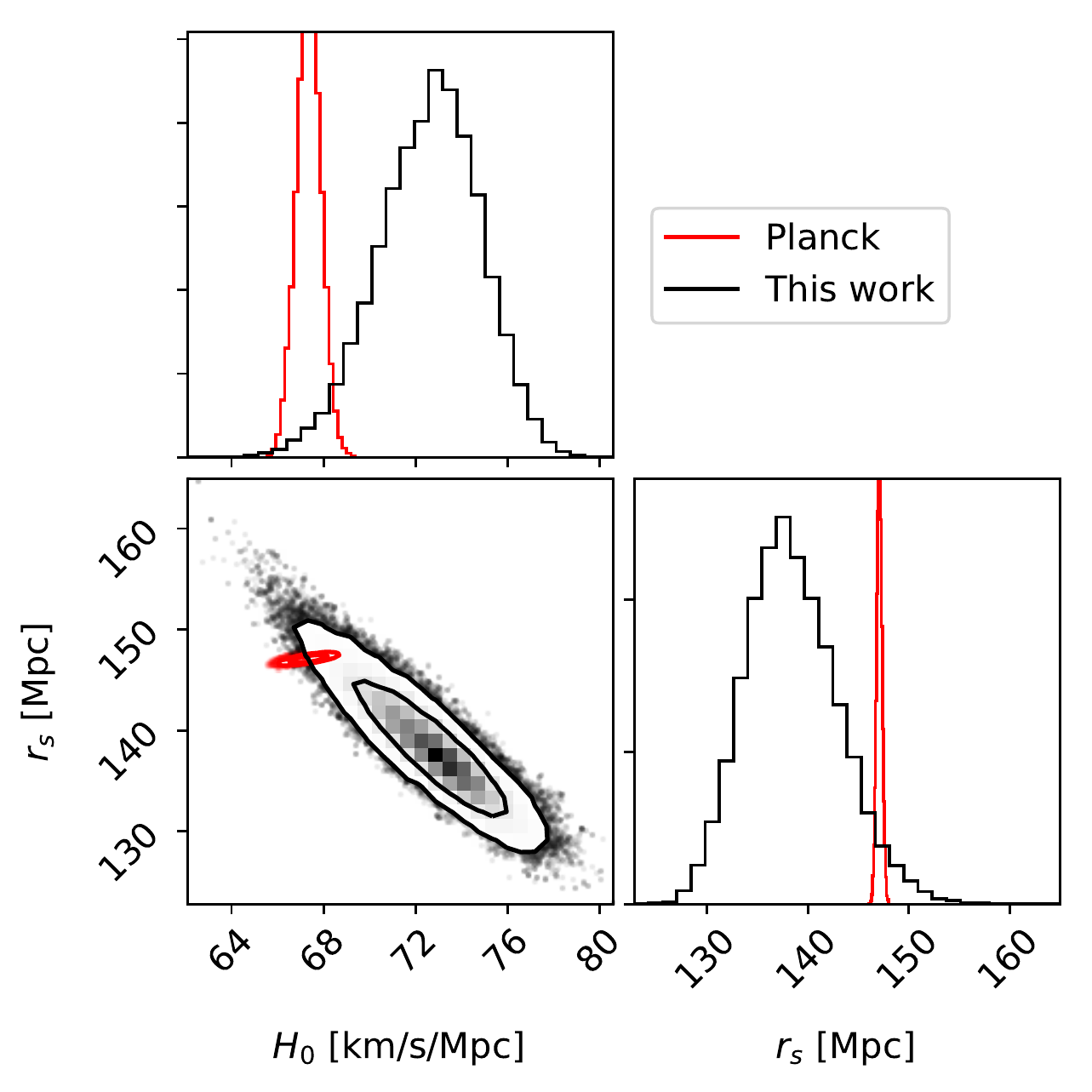}
        \caption{Inference on cosmological parameters, including the Hubble constant $H_0$ and sound horizon $r_s,$ for the baseline case of flat-$\Lambda$CDM models using time-delay lenses, SN~\textsc{I}a, and BAO as late-time indicators. {The outermost credibility contour contains 95\% of the marginalised posterior probability, and the innermost contour contains 68\%.}}
        \label{fig:lcdm}
\end{figure}

 \noindent The likelihoods of the lensed quasars HE0435, RXJ1131, and B1608 of the H0LiCOW collaboration were given as skewed log-normal distributions of their time-delay distances $D_{\Delta t}=(1+z_{l})D_{A,l}D_{A,s}/D_{A,ls}.$ For the lensed quasar J1206, both the angular diameter distance and the time delay distance were available in the form of a sample drawn from the model posterior distribution. A Gaussian kernel density estimator (KDE) was used to interpolate a smooth distribution between the posterior points.

The final log-likelihood that was sampled by \texttt{emcee} is a sum of the separate likelihoods of the SN, BAO, lensed quasars, and high-redshift quasars,
\begin{equation}
\ln \left( \mathcal{L}_{\textrm{total}} \right) = \ln (\mathcal{L}_{\textrm{SN}}) + \ln (\mathcal{L}_{\textrm{BAO}}) + \ln (\mathcal{L}_{\textrm{lenses}}) + \ln (\mathcal{L}_{\textrm{quasars}}).
\end{equation}
We note that the high-redshift quasar likelihood is optional in our study. For all cosmographic models used in our work, parameter inference is carried out with or without quasar data, and both results are consistently reported.

A uniform prior was used for all the free parameters, except when the high-redshift quasar sample was used. In that case, the intrinsic scatter $\sigma_{\textrm{int}}$ was also constrained to be larger than zero. This choice of priors does not seem to bias the inference according to current data and tests on flat-$\Lambda$CDM mocks.

To choose the right order of expansion for each model, the Bayesian information criterion (BIC) indicator was used,
\begin{equation}
    \mathrm{BIC}\ =\ \ln(N)k-2\ln(\mathcal{L}_{m.a.p.}),
\end{equation}
where $N$ is the number of data points, $k$ is the number of free parameters and $\mathcal{L}_{m.a.p.}$ is the \textit{\textup{maximum a posteriori}} likelihood (i.e. evaluated where the posterior is maximised). The BIC score expresses how well a model describes the data, with a lower score corresponding to a better agreement. It also introduces a penalty term for added complexity in a model. Table \ref{table:BICscores} displays the number of free parameters, maximum a posteriori likelihood, and the BIC score for four increasing orders of expansion. The expansion order with five free parameters provides the lowest BIC score. When more complexity is added to the model, the BIC value continuously increases, which supports the conclusion that higher expansion orders will be ruled out as well. When the high-redshift quasar sample was added to the data collection, it changed the preferred order of expansion of model 3 from third to second order.

\begin{table}[H]
\begin{tabular}{lllll}
\textbf{Model 1}                               &                                &                                &                                &           
\\
\multicolumn{1}{l|}{ }                      & \multicolumn{1}{l|}{ first $\; \; \; \; \;$} & \multicolumn{1}{l|}{ second $\; \; \; $} & \multicolumn{1}{l|}{ third $\; \; \; \; \; $} &   fourth $\; \; \; $
\\
\multicolumn{1}{l|}{parameter}                      & \multicolumn{1}{l|}{ order} & \multicolumn{1}{l|}{ order} & \multicolumn{1}{l|}{order} &   order \\ \hline
\multicolumn{1}{l|}{Free parameters $\; \; $}                       & \multicolumn{1}{l|}{4}         & \multicolumn{1}{l|}{5}         & \multicolumn{1}{l|}{6}         & 7         \\
\multicolumn{1}{l|}{$\ln\mathcal{L}_{m.a.p.}$} & \multicolumn{1}{l|}{-60.8}     & \multicolumn{1}{l|}{-55.8}     & \multicolumn{1}{l|}{-55.3}     & -55.2     \\
\multicolumn{1}{l|}{BIC score}                 & \multicolumn{1}{l|}{137.2}     & \multicolumn{1}{l|}{131.1}     & \multicolumn{1}{l|}{134.1}     & 137.8     \\
                                               &                                &                                &                                &           \\
\textbf{Model 2}                               &                                &                                &                                &           \\
\multicolumn{1}{l|}{ }                      & \multicolumn{1}{l|}{first} & \multicolumn{1}{l|}{second} & \multicolumn{1}{l|}{ third} &  fourth
\\
\multicolumn{1}{l|}{parameter}                      & \multicolumn{1}{l|}{order} & \multicolumn{1}{l|}{order} & \multicolumn{1}{l|}{ order} &  order \\ \hline
\multicolumn{1}{l|}{Free parameters}                       & \multicolumn{1}{l|}{4}         & \multicolumn{1}{l|}{5}         & \multicolumn{1}{l|}{6}         & 7         \\
\multicolumn{1}{l|}{$\ln\mathcal{L}_{m.a.p.}$} & \multicolumn{1}{l|}{-67.1}     & \multicolumn{1}{l|}{-56.8}     & \multicolumn{1}{l|}{-55.7}     & -55.0     \\
\multicolumn{1}{l|}{BIC score}                 & \multicolumn{1}{l|}{149.8}     & \multicolumn{1}{l|}{133.2}     & \multicolumn{1}{l|}{134.9}     & 137.4     \\
                                               &                                &                                &                                &           \\
\textbf{Model 3}                               &                                &                                &                                &           \\
\multicolumn{1}{l|}{ }                      & \multicolumn{1}{l|}{second} & \multicolumn{1}{l|}{third} & \multicolumn{1}{l|}{ fourth} &  fifth
\\
\multicolumn{1}{l|}{parameter}                      & \multicolumn{1}{l|}{order} & \multicolumn{1}{l|}{order} & \multicolumn{1}{l|}{ order} &  order \\ \hline
\multicolumn{1}{l|}{Free parameters}                       & \multicolumn{1}{l|}{4}         & \multicolumn{1}{l|}{5}         & \multicolumn{1}{l|}{6}         & 7         \\
\multicolumn{1}{l|}{$\ln\mathcal{L}_{m.a.p.}$} & \multicolumn{1}{l|}{-61.0}     & \multicolumn{1}{l|}{-56.1}     & \multicolumn{1}{l|}{-56.0}     & -54.5     \\
\multicolumn{1}{l|}{BIC score}                 & \multicolumn{1}{l|}{137.6}     & \multicolumn{1}{l|}{131.7}     & \multicolumn{1}{l|}{135.5}     & 136.3    \\
\hline \\

\end{tabular}
\caption{Overview of the number of free parameters, maximum a posteriori likelihood, and BIC score for different expansion orders for cosmographic models 1, 2, and 3. These numbers were calculated using the four lenses, SN, and BAO points and assuming a flat Universe. For expansion in H (models 1 and 2) the second order is preferred, and for expansion in distance (model 3) the third order is preferred. This corresponds to five free parameters in each of the models. }
\label{table:BICscores}
\end{table}

\begin{table*}
\centering          
\begin{tabular}{l l l l l}
\hline 

 & \multicolumn{4}{c}{flat ($\Omega_{k}=0$)} \\
parameter &  model~1 (second order) & model~2 (second order) & model~3 (third order) & model~4 (f$\Lambda$CDM)\\ 
\hline
\\
$r_{s}$ (Mpc) & $135.26 \pm 5.22$ & $138.38 \pm 4.97$ & $137.76 \pm 4.970$ & $138.74 \pm 4.67$\\
$H_{0}r_{s}$ (km~s$^{-1}$) & $10091.06 \pm  147.54$ & $ 10095.11\pm 146.23 $ & $10069.64\pm 149.82$ & $10046.10 \pm 137.33$\\
$H_{0}$ (km~s$^{-1}$~Mpc$^{-1}$) & $74.71 \pm 2.92$ & $73.06 \pm 2.65$ & $ 73.09 \pm 2.67$ & $72.48 \pm 2.24$\\
$q_{0}$ & $ - 0.62 \pm 0.078$ & $-0.72 \pm 0.11$ & $ - 0.57 \pm 0.18$ & ---\\
$\ln\mathcal{L}_{m.a.p.}$ & $-55.76$ & $-56.80$ & $-56.06$ & $-56.31$ \\
BIC score & 131.07 & 133.15 & 131.68 & 128.30\\
$\ln\tau$ (Planck $\Lambda$CDM) &3.1 (2.0$\sigma$) &2.3 (1.6$\sigma$) &2.3 (1.7$\sigma$) & 2.5 (1.8$\sigma$)\\
\hline\hline
 & \multicolumn{4}{c}{free $\Omega_{k}$} \\
parameter &  model~1 (second order) & model~2 (second order) & model~3 (third order) & model~4 ($\Lambda$CDM)\\ 
\hline
\\
$r_{s}$ (Mpc) & $133.04\pm 7.57$ & $137.57 \pm 7.80$ & $136.19 \pm 8.05$ & $139.91 \pm 5.54$\\
$H_{0}r_{s}$ (km~s$^{-1}$) & $10069.16 \pm 156.97$ & $10079.25 \pm 158.20$ & $10052.22 \pm 162.32 $ & $ 10073.39 \pm 155.18$\\
$H_{0}$ (km~s$^{-1}$~Mpc$^{-1}$) & $75.91 \pm 4.07$ & $73.48\pm 3.86$ & $73.82 \pm 4.06$ & $72.09 \pm 2.41$\\
$\Omega_{k}$ & $0.099 \pm 0.23$ & $ 0.038 \pm 0.21$ & $ 0.079\pm 0.22$ & $-0.066 \pm 0.16$ \\
$q_{0}$ & $-0.62 \pm 0.087$ & $- 0.71 \pm 0.11$ & $ -0.55 \pm 0.23$ & ---\\
$\ln\mathcal{L}_{m.a.p.}$ &  $-56.08$ & $-57.10$ & $-56.32$ & $-56.19$ \\
BIC score & 135.63 & 137.67 & 136.11 & 131.95\\
$\ln\tau$ (Planck $\Lambda$CDM) & 2.3 (1.6$\sigma$) & 1.6 (1.3$\sigma$) & 1.5 (1.2$\sigma$) & 2.3 (1.6 $\sigma$) \\
\hline
\end{tabular}
\caption{Inference on the cosmological parameters from BAO+SNe+lenses in our four model classes, with or without imposed flatness. {We list the posterior mean and 68\% uncertainties of the main parameters, the maximum a posteriori likelihood, the BIC score, and the odds $\tau$ that our measurements of $H_{0}$ and $r_{\rm s}$ are consistent with those from the \textit{Planck} observations, as derived for the standard flat-$\Lambda$CDM cosmological model.}}
\label{table:results}  
\end{table*}
\begin{table*}
\centering          
\begin{tabular}{l l l l l}
\hline 

 & \multicolumn{4}{c}{flat ($\Omega_{k}=0$)} \\
parameter &  model~1 (second order) & model~2 (second order) & model~3 (second order) & model~4 (f$\Lambda$CDM)\\ 
\hline
\\
$r_{s}$ (Mpc) & $132.36 \pm 5.05$ & $135.67 \pm 4.84$ & $131.63 \pm 4.45$ & $138.24 \pm 4.62$\\
$H_{0}r_{s}$ (km~s$^{-1}$) & $10124.73\pm 143.40$ & $10111.40 \pm 147.68 $ & $10186.40 \pm 145.68$ & $9999.72 \pm 134.38$\\
$H_{0}$ (km~s$^{-1}$~Mpc$^{-1}$) & $76.59 \pm 2.90$ & $74.62 \pm 2.67 $ & $77.38 \pm 2.52$ & $72.40 \pm 2.21$\\
$q_{0}$ & $- 0.70 \pm 0.074$ & $ -0.82 \pm 0.105$ & $ -1.13 \pm 0.11$ & ---\\
$\ln\mathcal{L}_{m.a.p.}$ & $-2335.33$ & $-2338.02$ & $-2339.59$ & $-2338.14$ \\
BIC score & 4720.84 & 4726.22 & 4722.19 & 4719.30\\
$\ln\tau$ (Planck $\Lambda$CDM) & 4.9 (2.7$\sigma$) & 3.5 (2.2$\sigma$) &7.8 (3.5$\sigma$)& 2.5 (1.7$\sigma$)\\
\hline\hline
 & \multicolumn{4}{c}{free $\Omega_{k}$} \\
parameter &  model~1 (second order) & model~2 (second order) & model~3 (second order) & model~4 ($\Lambda$CDM)\\ 
\hline
\\
$r_{s}$ (Mpc) & $ 134.20  \pm 8.00 $ & $ 140.74 \pm 8.15$ & $ 139.36 \pm 8.40$ & $ 143.70 \pm 5.58$\\
$H_{0}r_{s}$ (km~s$^{-1}$) & $10132.11 \pm 160.61$ & $ 10150.20 \pm 155.94$ & $10223.94 \pm 152.08$ & $ 10140.36 \pm 157.6$\\
$H_{0}$ (km~s$^{-1}$~Mpc$^{-1}$) & $ 75.74 \pm 4.16$ & $ 72.34 \pm 3.89$ & $ 73.37 \pm 4.18$ & $ 70.65 \pm 2.29$\\
$\Omega_{k}$ & $ - 0.056 \pm 0.22$ & $ -0.16 \pm 0.20$ & $  -0.19 \pm 0.17$ &  $ -0.27 \pm 0.14$\\
$q_{0}$ & $ -0.70 \pm 0.082$ & $ -0.82 \pm 0.11$ & $ -1.11 \pm 0.17$ & ---\\
$\ln\mathcal{L}_{m.a.p.}$ &  $-2335.58$ & $-2337.84$ & $-2339.30$ & $-2336.20$ \\
BIC score & 4728.53 & 4733.03 & 4728.76 & 4722.56 \\
$\ln\tau$ (Planck $\Lambda$CDM) & 2.4 (1.7$\sigma$) & 0.6 (0.6$\sigma$) & 2.6 (1.8$\sigma$) & 1.9 (1.4$\sigma$) \\
\hline
\end{tabular}
\caption{Same as for Table~1, but including UV-Xray quasars as standardisable distance indicators.}
\label{table:withquasars}   
\end{table*}

\begin{figure}
        \centering
        \includegraphics[width=\linewidth]{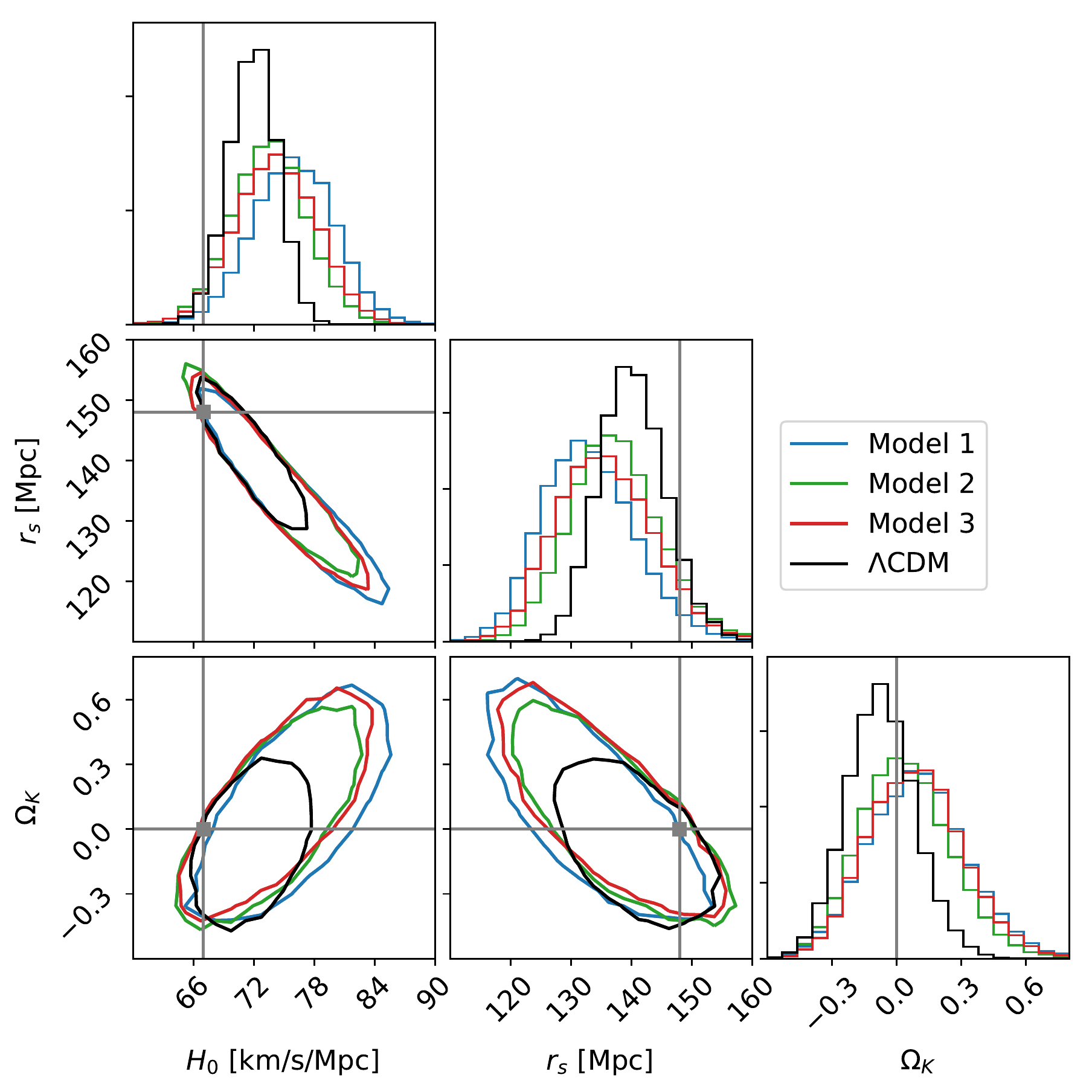}
        \caption{Inference on the Hubble constant $H_0$ and sound horizon $r_s$ for different models (at fiducial truncation order for models~1-3), with free $\Omega_{k},$ using time-delay lenses, SN~\textsc{I}a, and BAO. While the inferred parameters can change among models and among truncation choices, the relative discrepancy with CMB measurements remains the same. {The credibility contours contain 95\% of the marginalised posterior probability. The grey point corresponds to the \textit{Planck} value of $H_0$ and $r_s$ and to a flat Universe.}}
        \label{fig:4models}
\end{figure}

\section{Results and discussion}

The inferred values from our inference are given in Tables \ref{table:results} and \ref{table:withquasars}. For the sake of compactness, we report only the inferred values for each model that correspond to the lowest BIC scores (and to a $\Delta$BIC > 2). Figure \ref{fig:trends} shows the change in $H_0$ as inferred by different expansion orders. Plots of marginalised posteriors on selected cosmological parameters are given in Figures \ref{fig:lcdm} and \ref{fig:4models}. 

The inferred values of the Hubble constant from Table 1, {both its maximum a posteriori and uncertainty, vary between $(73.0 \pm 2.7)$~km~s$^{-1}$~Mpc$^{-1}$  and $(76.0 \pm 4.0)$~km~s$^{-1}$~Mpc$^{-1}$.} They are in full agreement with current results form the H0LiCOW and SH0ES collaborations, even despite the choice of general and agnostic models in our method. This indicates that the discrepancy between Cepheid-calibrated $H_0$ and that inferred from CMB measurements is not due to (known and unknown) systematics in the very low redshift range. The inferred sound horizon $r_s$ {varies between $(133 \pm 8)$~Mpc  and $(138 \pm 5)$~Mpc. }
The largest discrepancy with the value from CMB and Standard Model predictions ($147.09\pm0.26$~Mpc) is more significant for models that are agnostic to the underlying cosmology.

The systematic uncertainties, due to different model choices, are still within the range allowed by statistical uncertainties. However, they may become dominant in future measurements aiming at percent-level precision. Adding UV-Xray standardisable quasars generally raises the inferred value of $H_0$ (and correspondingly lowers the inferred $r_s$), even though the normalisation of their Hubble diagram is treated as a nuisance parameter. The addition of the quasar sample also results in lower values of $\Omega_k$. This suggests that the behaviour of distance modulus with redshift has sufficient constraining power on auxiliary cosmological parameters that  in turn are degenerate with $H_0$ in the time-delay lensing standardisation. For all cosmographic models, the intrinsic scatter in quasar distance moduli found in our analysis is 1.45 mag, which is fully consistent with the estimate reported in \cite{Ris2018}.

We quantified the tension with CMB measurements through the two-dimensional inference on $H_0$ and $r_s.$ Following \citet{Ver2013}, we estimated the odds that both measurements are consistent by computing the following ratio:
\begin{equation}
    \tau = \frac{\int\int \hat{p}_{\rm CMB}\hat{p}_{\rm local} \textrm{d}H_{0}\textrm{d}r_{\rm s}}{\int\int p_{\rm CMB}p_{\rm local} \textrm{d}H_{0}\textrm{d}r_{\rm s}}
,\end{equation}
where $p$ is the marginalised probability distribution for $r_{\rm s}$ and $H_{0}$ from the CMB \citep{Planck2018} or our study (in both cases approximated by Gaussians), while $\hat{p}$ is a distribution shifted to a fixed arbitrary point so that both measurements have the same posterior probability means. A more intuitive scale representing the discrepancy between two measurements is a number-of-sigma tension, which can be derived from the odds ratio. This is done by calculating the probability enclosed by a contour such that $1 / \tau = e^{- \frac{1}{2} r^2} $. The number of sigma tension can then be calculated from the probability by means of the error function. We list the logarithm of the odds and the number of sigma tension in Tables \ref{table:results} and \ref{table:withquasars}.

The tension with Planck measurements from CMB is approximately at a $2\sigma$ level. While the uncertainties from some model families are larger, the corresponding $H_0$ ($r_s$) optimal values are also higher (lower), and the tension remains the same.  {The curvature $\Omega_k$ slightly alleviates the tension through larger $H_0$ uncertainties, but the current data do not yield any evidence of a departure from flatness.}

\section{Conclusions and outlook}
Current data enable a $\approx3\%$ determination of key cosmological parameters, in particular, the Hubble constant $H_0$ and the sound horizon $r_s,$ resulting in a $\approx 2\sigma$ Gaussian tension with predictions from CMB measurements and the Standard Model. While this tension is robust against the choice of model family and is therefore independent of the underlying cosmology, the systematics due to different model choices are currently comparable to the statistical uncertainties and may dominate percent-level measurements of $H_0.$ A simple estimate based on recent SH0ES measurements \citep{Riess2019} and very recent five-lens measurements by H0LiCOW \citep{Rusu2019} indicates a $\approx5\sigma$ tension with CMB measurements within a flat-$\Lambda$CDM model.

Our study also demonstrated the potential of constraining the curvature of the Universe solely based on low-redshift observations and in a cosmology-independent manner. The current precision of $0.20$ is insufficient to test possible minimum departures from flatness, mainly due to the accuracy in $H_0$ from a small sample of well-studied lenses. Samples of lenses with suitable ancillary data are already being assembled \citep[see e.g.][]{sha19}. Future measurements of gravitational time-delays from the Large Synoptic Survey Telescope can reach percent-level precision \citep{kai15}, making this method a highly competitive probe \citep{Den2018}. 

\begin{acknowledgements}
 The authors were supported by a grant from VILLUM FONDEN (project number 16599). This project is partially funded by the Danish Council for independent research under the project ``Fundamentals of Dark Matter Structures'', DFF--6108-00470.\\
 We are grateful for the public release of the time-delay distance likelihoods by the H0LiCOW collaboration, and interesting conversations with S.~H. Suyu, F. Courbin and T. Treu.\\
We thank Guido Risaliti for sharing distance moduli measured from high-redshift quasars. \\
 We also thank the anonymous referee for constructive comments which helped improve our work.
 
\end{acknowledgements}

 
\bibliography{main}

\begin{thebibliography}{32}
\expandafter\ifx\csname natexlab\endcsname\relax\def\natexlab#1{#1}\fi

\bibitem[{{Alam} {et~al.}(2017){Alam}, {Ata}, {Bailey}, {Beutler}, {Bizyaev},
  {Blazek}, {Bolton}, {Brownstein}, {Burden}, {Chuang}, {Comparat}, {Cuesta},
  {Dawson}, {Eisenstein}, {Escoffier}, {Gil-Mar{\'\i}n}, {Grieb}, {Hand}, {Ho},
  {Kinemuchi}, {Kirkby}, {Kitaura}, {Malanushenko}, {Malanushenko}, {Maraston},
  {McBride}, {Nichol}, {Olmstead}, {Oravetz}, {Padmanabhan},
  {Palanque-Delabrouille}, {Pan}, {Pellejero-Ibanez}, {Percival}, {Petitjean},
  {Prada}, {Price-Whelan}, {Reid}, {Rodr{\'\i}guez-Torres}, {Roe}, {Ross},
  {Ross}, {Rossi}, {Rubi{\~n}o-Mart{\'\i}n}, {Saito}, {Salazar-Albornoz},
  {Samushia}, {S{\'a}nchez}, {Satpathy}, {Schlegel}, {Schneider},
  {Sc{\'o}ccola}, {Seo}, {Sheldon}, {Simmons}, {Slosar}, {Strauss}, {Swanson},
  {Thomas}, {Tinker}, {Tojeiro}, {Maga{\~n}a}, {Vazquez}, {Verde}, {Wake},
  {Wang}, {Weinberg}, {White}, {Wood-Vasey}, {Y{\`e}che}, {Zehavi}, {Zhai}, \&
  {Zhao}}]{Alam2017}
{Alam}, S., {Ata}, M., {Bailey}, S., {et~al.} 2017, \mnras, 470, 2617

\bibitem[{{Aylor} {et~al.}(2019){Aylor}, {Joy}, {Knox}, {Millea},
  {Raghunathan}, \& {Kimmy Wu}}]{Ayl2019}
{Aylor}, K., {Joy}, M., {Knox}, L., {et~al.} 2019, \apj, 874, 4

\bibitem[{{Bernal} {et~al.}(2016){Bernal}, {Verde}, \& {Riess}}]{Ber2016}
{Bernal}, J.~L., {Verde}, L., \& {Riess}, A.~G. 2016, \jcap, 10, 019

\bibitem[{{Betoule} {et~al.}(2014){Betoule}, {Kessler}, {Guy}, {Mosher},
  {Hardin}, {Biswas}, {Astier}, {El-Hage}, {Konig}, {Kuhlmann}, {Marriner},
  {Pain}, {Regnault}, {Balland}, {Bassett}, {Brown}, {Campbell}, {Carlberg},
  {Cellier-Holzem}, {Cinabro}, {Conley}, {D'Andrea}, {DePoy}, {Doi}, {Ellis},
  {Fabbro}, {Filippenko}, {Foley}, {Frieman}, {Fouchez}, {Galbany}, {Goobar},
  {Gupta}, {Hill}, {Hlozek}, {Hogan}, {Hook}, {Howell}, {Jha}, {Le Guillou},
  {Leloudas}, {Lidman}, {Marshall}, {M{\"o}ller}, {Mour{\~a}o}, {Neveu},
  {Nichol}, {Olmstead}, {Palanque-Delabrouille}, {Perlmutter}, {Prieto},
  {Pritchet}, {Richmond}, {Riess}, {Ruhlmann-Kleider}, {Sako}, {Schahmaneche},
  {Schneider}, {Smith}, {Sollerman}, {Sullivan}, {Walton}, \&
  {Wheeler}}]{Bet2014}
{Betoule}, M., {Kessler}, R., {Guy}, J., {et~al.} 2014, \aap, 568, A22

\bibitem[{{Birrer} {et~al.}(2019){Birrer}, {Treu}, {Rusu}, {Bonvin},
  {Fassnacht}, {Chan}, {Agnello}, {Shajib}, {Chen}, {Auger}, {Courbin},
  {Hilbert}, {Sluse}, {Suyu}, {Wong}, {Marshall}, {Lemaux}, \&
  {Meylan}}]{Bir2019}
{Birrer}, S., {Treu}, T., {Rusu}, C.~E., {et~al.} 2019, \mnras, 484, 4726

\bibitem[{{Denissenya} {et~al.}(2018){Denissenya}, {Linder}, \&
  {Shafieloo}}]{Den2018}
{Denissenya}, M., {Linder}, E.~V., \& {Shafieloo}, A. 2018, \jcap, 3, 041

\bibitem[{{Foreman-Mackey} {et~al.}(2013){Foreman-Mackey}, {Hogg}, {Lang}, \&
  {Goodman}}]{fm13}
{Foreman-Mackey}, D., {Hogg}, D.~W., {Lang}, D., \& {Goodman}, J. 2013,
  Publications of the Astronomical Society of the Pacific, 125, 306

\bibitem[{{Goodman} \& {Weare}(2010)}]{gw2010}
{Goodman}, J. \& {Weare}, J. 2010, Communications in Applied Mathematics and
  Computational Science, Vol.~5, No.~1, p.~65-80, 2010, 5, 65

\bibitem[{{Heavens} {et~al.}(2014){Heavens}, {Jimenez}, \& {Verde}}]{Hea2014}
{Heavens}, A., {Jimenez}, R., \& {Verde}, L. 2014, Physical Review Letters,
  113, 241302

\bibitem[{{Jee} {et~al.}(2016){Jee}, {Komatsu}, {Suyu}, \& {Huterer}}]{Jee2016}
{Jee}, I., {Komatsu}, E., {Suyu}, S.~H., \& {Huterer}, D. 2016, \jcap, 2016,
  031

\bibitem[{{L'Huillier} \& {Shafieloo}(2017)}]{Hui2017}
{L'Huillier}, B. \& {Shafieloo}, A. 2017, Journal of Cosmology and
  Astro-Particle Physics, 2017, 015

\bibitem[{{Li} {et~al.}(2019){Li}, {Du}, \& {Xu}}]{Li2019}
{Li}, E.-K., {Du}, M., \& {Xu}, L. 2019, arXiv e-prints, arXiv:1903.11433

\bibitem[{{Liao} {et~al.}(2015){Liao}, {Treu}, {Marshall}, {Fassnacht},
  {Rumbaugh}, {Dobler}, {Aghamousa}, {Bonvin}, {Courbin}, {Hojjati}, {Jackson},
  {Kashyap}, {Rathna Kumar}, {Linder}, {Mandel}, {Meng}, {Meylan}, {Moustakas},
  {Prabhu}, {Romero-Wolf}, {Shafieloo}, {Siemiginowska}, {Stalin}, {Tak},
  {Tewes}, \& {van Dyk}}]{kai15}
{Liao}, K., {Treu}, T., {Marshall}, P., {et~al.} 2015, \apj, 800, 11

\bibitem[{{Macaulay} {et~al.}(2019){Macaulay}, {Nichol}, {Bacon}, {Brout},
  {Davis}, {Zhang}, {Bassett}, {Scolnic}, {M{\"o}ller}, {D'Andrea}, {Hinton},
  {Kessler}, {Kim}, {Lasker}, {Lidman}, {Sako}, {Smith}, {Sullivan}, {Abbott},
  {Allam}, {Annis}, {Asorey}, {Avila}, {Bechtol}, {Brooks}, {Brown}, {Burke},
  {Calcino}, {Carnero Rosell}, {Carollo}, {Carrasco Kind}, {Carretero},
  {Castander}, {Collett}, {Crocce}, {Cunha}, {da Costa}, {Davis}, {De Vicente},
  {Diehl}, {Doel}, {Drlica-Wagner}, {Eifler}, {Estrada}, {Evrard},
  {Filippenko}, {Finley}, {Flaugher}, {Foley}, {Fosalba}, {Frieman}, {Galbany},
  {Garc{\'{\i}}a-Bellido}, {Gaztanaga}, {Glazebrook},
  {Gonz{\'a}lez-Gait{\'a}n}, {Gruen}, {Gruendl}, {Gschwend}, {Gutierrez},
  {Hartley}, {Hollowood}, {Honscheid}, {Hoormann}, {Hoyle}, {Huterer}, {Jain},
  {James}, {Jeltema}, {Kasai}, {Krause}, {Kuehn}, {Kuropatkin}, {Lahav},
  {Lewis}, {Li}, {Lima}, {Lin}, {Maia}, {Marshall}, {Martini}, {Miquel},
  {Nugent}, {Palmese}, {Pan}, {Plazas}, {Romer}, {Roodman}, {Sanchez},
  {Scarpine}, {Schindler}, {Schubnell}, {Serrano}, {Sevilla-Noarbe}, {Sharp},
  {Soares-Santos}, {Sobreira}, {Sommer}, {Suchyta}, {Swann}, {Swanson},
  {Tarle}, {Thomas}, {Thomas}, {Tucker}, {Uddin}, {Vikram}, {Walker}, \&
  {Wiseman}}]{mac19}
{Macaulay}, E., {Nichol}, R.~C., {Bacon}, D., {et~al.} 2019, \mnras, 486, 2184

\bibitem[{{Planck Collaboration} {et~al.}(2018){Planck Collaboration},
  {Aghanim}, {Akrami}, {Ashdown}, {Aumont}, {Baccigalupi}, {Ballardini},
  {Banday}, {Barreiro}, {Bartolo}, {Basak}, {Battye}, {Benabed}, {Bernard},
  {Bersanelli}, {Bielewicz}, {Bock}, {Bond}, {Borrill}, {Bouchet}, {Boulanger},
  {Bucher}, {Burigana}, {Butler}, {Calabrese}, {Cardoso}, {Carron},
  {Challinor}, {Chiang}, {Chluba}, {Colombo}, {Combet}, {Contreras}, {Crill},
  {Cuttaia}, {de Bernardis}, {de Zotti}, {Delabrouille}, {Delouis}, {Di
  Valentino}, {Diego}, {Dor{\'e}}, {Douspis}, {Ducout}, {Dupac}, {Dusini},
  {Efstathiou}, {Elsner}, {En{\ss}lin}, {Eriksen}, {Fantaye}, {Farhang},
  {Fergusson}, {Fernandez-Cobos}, {Finelli}, {Forastieri}, {Frailis},
  {Franceschi}, {Frolov}, {Galeotta}, {Galli}, {Ganga}, {G{\'e}nova-Santos},
  {Gerbino}, {Ghosh}, {Gonz{\'a}lez-Nuevo}, {G{\'o}rski}, {Gratton},
  {Gruppuso}, {Gudmundsson}, {Hamann}, {Hand ley}, {Herranz}, {Hivon}, {Huang},
  {Jaffe}, {Jones}, {Karakci}, {Keih{\"a}nen}, {Keskitalo}, {Kiiveri}, {Kim},
  {Kisner}, {Knox}, {Krachmalnicoff}, {Kunz}, {Kurki-Suonio}, {Lagache},
  {Lamarre}, {Lasenby}, {Lattanzi}, {Lawrence}, {Le Jeune}, {Lemos},
  {Lesgourgues}, {Levrier}, {Lewis}, {Liguori}, {Lilje}, {Lilley}, {Lindholm},
  {L{\'o}pez-Caniego}, {Lubin}, {Ma}, {Mac{\'\i}as-P{\'e}rez}, {Maggio},
  {Maino}, {Mandolesi}, {Mangilli}, {Marcos-Caballero}, {Maris}, {Martin},
  {Martinelli}, {Mart{\'\i}nez-Gonz{\'a}lez}, {Matarrese}, {Mauri}, {McEwen},
  {Meinhold}, {Melchiorri}, {Mennella}, {Migliaccio}, {Millea}, {Mitra},
  {Miville-Desch{\^e}nes}, {Molinari}, {Montier}, {Morgante}, {Moss}, {Natoli},
  {N{\o}rgaard-Nielsen}, {Pagano}, {Paoletti}, {Partridge}, {Patanchon},
  {Peiris}, {Perrotta}, {Pettorino}, {Piacentini}, {Polastri}, {Polenta},
  {Puget}, {Rachen}, {Reinecke}, {Remazeilles}, {Renzi}, {Rocha}, {Rosset},
  {Roudier}, {Rubi{\~n}o-Mart{\'\i}n}, {Ruiz-Granados}, {Salvati}, {Sandri},
  {Savelainen}, {Scott}, {Shellard}, {Sirignano}, {Sirri}, {Spencer},
  {Sunyaev}, {Suur-Uski}, {Tauber}, {Tavagnacco}, {Tenti}, {Toffolatti},
  {Tomasi}, {Trombetti}, {Valenziano}, {Valiviita}, {Van Tent}, {Vibert},
  {Vielva}, {Villa}, {Vittorio}, {Wand elt}, {Wehus}, {White}, {White},
  {Zacchei}, \& {Zonca}}]{Planck2018}
{Planck Collaboration}, {Aghanim}, N., {Akrami}, Y., {et~al.} 2018, arXiv
  e-prints, arXiv:1807.06209

\bibitem[{{Riess} {et~al.}(2019){Riess}, {Casertano}, {Yuan}, {Macri}, \&
  {Scolnic}}]{Riess2019}
{Riess}, A.~G., {Casertano}, S., {Yuan}, W., {Macri}, L.~M., \& {Scolnic}, D.
  2019, \apj, 876, 85

\bibitem[{{Risaliti} \& {Lusso}(2018)}]{Ris2018}
{Risaliti}, G. \& {Lusso}, E. 2018, arXiv e-prints, arXiv:1811.02590

\bibitem[{{Rusu} {et~al.}(2019){Rusu}, {Wong}, {Bonvin}, {Sluse}, {Suyu},
  {Fassnacht}, {Chan}, {Hilbert}, {Auger}, {Sonnenfeld}, {Birrer}, {Courbin},
  {Treu}, {Chen}, {Halkola}, {Koopmans}, {Marshall}, \& {Shajib}}]{Rusu2019}
{Rusu}, C.~E., {Wong}, K.~C., {Bonvin}, V., {et~al.} 2019, arXiv e-prints
  [\eprint[arXiv]{1905.09338}]

\bibitem[{{Scolnic} {et~al.}(2018){Scolnic}, {Jones}, {Rest}, {Pan},
  {Chornock}, {Foley}, {Huber}, {Kessler}, {Narayan}, {Riess}, {Rodney},
  {Berger}, {Brout}, {Challis}, {Drout}, {Finkbeiner}, {Lunnan}, {Kirshner},
  {Sand ers}, {Schlafly}, {Smartt}, {Stubbs}, {Tonry}, {Wood-Vasey}, {Foley},
  {Hand}, {Johnson}, {Burgett}, {Chambers}, {Draper}, {Hodapp}, {Kaiser},
  {Kudritzki}, {Magnier}, {Metcalfe}, {Bresolin}, {Gall}, {Kotak}, {McCrum}, \&
  {Smith}}]{Scol2018}
{Scolnic}, D.~M., {Jones}, D.~O., {Rest}, A., {et~al.} 2018, \apj, 859, 101

\bibitem[{{Shafieloo} {et~al.}(2018){Shafieloo}, {L'Huillier}, \&
  {Starobinsky}}]{Sha2018}
{Shafieloo}, A., {L'Huillier}, B., \& {Starobinsky}, A.~A. 2018, \prd, 98,
  083526

\bibitem[{{Shajib} {et~al.}(2019){Shajib}, {Birrer}, {Treu}, {Auger},
  {Agnello}, {Anguita}, {Buckley-Geer}, {Chan}, {Collett}, {Courbin},
  {Fassnacht}, {Frieman}, {Kayo}, {Lemon}, {Lin}, {Marshall}, {McMahon},
  {More}, {Morgan}, {Motta}, {Oguri}, {Ostrovski}, {Rusu}, {Schechter},
  {Shanks}, {Suyu}, {Meylan}, {Abbott}, {Allam}, {Annis}, {Avila}, {Bertin},
  {Brooks}, {Carnero Rosell}, {Carrasco Kind}, {Carretero}, {Cunha}, {da
  Costa}, {De Vicente}, {Desai}, {Doel}, {Flaugher}, {Fosalba},
  {Garc{\'{\i}}a-Bellido}, {Gerdes}, {Gruen}, {Gruendl}, {Gutierrez},
  {Hartley}, {Hollowood}, {Hoyle}, {James}, {Kuehn}, {Kuropatkin}, {Lahav},
  {Lima}, {Maia}, {March}, {Marshall}, {Melchior}, {Menanteau}, {Miquel},
  {Plazas}, {Sanchez}, {Scarpine}, {Sevilla-Noarbe}, {Smith}, {Soares-Santos},
  {Sobreira}, {Suchyta}, {Swanson}, {Tarle}, \& {Walker}}]{sha19}
{Shajib}, A.~J., {Birrer}, S., {Treu}, T., {et~al.} 2019, \mnras, 483, 5649

\bibitem[{{Suyu} {et~al.}(2017){Suyu}, {Bonvin}, {Courbin}, {Fassnacht},
  {Rusu}, {Sluse}, {Treu}, {Wong}, {Auger}, {Ding}, {Hilbert}, {Marshall},
  {Rumbaugh}, {Sonnenfeld}, {Tewes}, {Tihhonova}, {Agnello}, {Blandford},
  {Chen}, {Collett}, {Koopmans}, {Liao}, {Meylan}, \& {Spiniello}}]{Suyu2017}
{Suyu}, S.~H., {Bonvin}, V., {Courbin}, F., {et~al.} 2017, \mnras, 468, 2590

\bibitem[{{Suyu} {et~al.}(2010){Suyu}, {Marshall}, {Auger}, {Hilbert},
  {Blandford}, {Koopmans}, {Fassnacht}, \& {Treu}}]{Suyu2010}
{Suyu}, S.~H., {Marshall}, P.~J., {Auger}, M.~W., {et~al.} 2010, \apj, 711, 201

\bibitem[{{Suyu} {et~al.}(2014){Suyu}, {Treu}, {Hilbert}, {Sonnenfeld},
  {Auger}, {Blandford}, {Collett}, {Courbin}, {Fassnacht}, {Koopmans},
  {Marshall}, {Meylan}, {Spiniello}, \& {Tewes}}]{Suyu2014}
{Suyu}, S.~H., {Treu}, T., {Hilbert}, S., {et~al.} 2014, \apjl, 788, L35

\bibitem[{{Taubenberger} {et~al.}(2019){Taubenberger}, {Suyu}, {Komatsu},
  {Jee}, {Birrer}, {Bonvin}, {Courbin}, {Rusu}, {Shajib}, \& {Wong}}]{Taub2019}
{Taubenberger}, S., {Suyu}, S.~H., {Komatsu}, E., {et~al.} 2019, \aap, 628, L7

\bibitem[{{Verde} {et~al.}(2017){Verde}, {Bernal}, {Heavens}, \&
  {Jimenez}}]{Ver2017}
{Verde}, L., {Bernal}, J.~L., {Heavens}, A.~F., \& {Jimenez}, R. 2017, \mnras,
  467, 731

\bibitem[{{Verde} {et~al.}(2013){Verde}, {Protopapas}, \& {Jimenez}}]{Ver2013}
{Verde}, L., {Protopapas}, P., \& {Jimenez}, R. 2013, Physics of the Dark
  Universe, 2, 166

\bibitem[{{Visser}(2004)}]{Visser2004}
{Visser}, M. 2004, Classical and Quantum Gravity, 21, 2603

\bibitem[{{Weinberg}(2008)}]{Weinberg2008}
{Weinberg}, S. 2008, {Cosmology}

\bibitem[{{Wojtak} \& {Agnello}(2019)}]{Woj2019}
{Wojtak}, R. \& {Agnello}, A. 2019, \mnras, 486, 5046

\bibitem[{{Wong} {et~al.}(2017){Wong}, {Suyu}, {Auger}, {Bonvin}, {Courbin},
  {Fassnacht}, {Halkola}, {Rusu}, {Sluse}, {Sonnenfeld}, {Treu}, {Collett},
  {Hilbert}, {Koopmans}, {Marshall}, \& {Rumbaugh}}]{Wong2017}
{Wong}, K.~C., {Suyu}, S.~H., {Auger}, M.~W., {et~al.} 2017, \mnras, 465, 4895

\bibitem[{{Xu} \& {Wang}(2011)}]{Xu2011}
{Xu}, L. \& {Wang}, Y. 2011, Physics Letters B, 702, 114

\end{thebibliography}

\end{document}